# Supersymetric laser arrays

R. El-Ganainy[*]

*Department of Physics, Michigan Technological University, Houghton, Michigan 49931, USA*

Li Ge

*Department of Engineering Science and Physics, College of Staten Island, CUNY, Staten Island, NY 10314, USA*
*The Graduate Center, CUNY, New York, NY 10016, USA*

M. M. Khajavikhan and D.N. Christodoulides

*College of Optics & Photonics-CREOL, University of Central Florida, Orlando, Florida, USA*

**Abstract**

We introduce the concept of supersymmetric laser arrays that consists of a main optical lattice and its superpartner structure, and we investigate the onset of their lasing oscillations. Due to the coupling of the two constituent lattices, their degenerate optical modes form doublets, while the extra mode associated with unbroken supersymmetry forms a singlet state. Singlet lasing can be achieved for a wide range of design parameters either by introducing stronger loss in the partner lattice or by pumping only the main array. Our findings suggest the possibility of building single-mode, high-power laser arrays and are also important for understanding light transport dynamics in multimode Parity-Time symmetric photonic structures.

# 1. INTRODUCTION

Supersymmetry (SUSY) was proposed as a unifying theme that treats bosonic and fermionic particles on equal foot [1]. Later, this notion was applied to quantum mechanics, scattering processes and nonlinear dynamics [2]. By noting that SUSY transformations are not pertinent to quantum field theory, similar concepts were also applied to quantum cascaded lasers [3]. Recently, it was recognized that SUSY can be employed to achieve unprecedented control over light transport in optical guiding geometries [4] where mode conversion in passive SUSY optical waveguide arrays has been experimentally demonstrated [5].

Another field that has received much attention recently is pump-induced mode selection in integrated laser systems [6-10]. In these works, it was shown that laser emission of certain modes can be enhanced or suppressed by using localized pumping profiles. Noteworthy, the operation of these devices relies on the whereabouts of the so called exceptional points (EPs). These points are non-Hermitian degeneracies that occur when two or more eigenvalues and their corresponding eigenvectors coalesce [11]. Mathematically, they represent algebraic branch cut singularities at which the eigenvector space ceases to be complete [11, 12]. Recently, it was shown that lasing near EPs can lead to laser self-termination in coupled photonic molecules [13, 14]. Interestingly, it was found that this effect can be fully understood by using coupled mode formalism [15]. The work in Ref. [15] also provides a means for understanding and controlling the lasing properties of photonic molecules made of multiple photonic cavities. More recently, the concept of Parity-Time reversal (*PT*) symmetry was also invoked to build single longitudinal mode mircoring laser systems [16, 17].

In this context, it is interesting to note that the problem of laser oscillations in coupled photonic structures dates back to few decades ago. In fact, waveguide laser arrays have been a subject of intense investigations for the purpose of building high power phase-locked lasers [18-21]. However, it was shown that their operation is dominated by multimode chaotic emission [19]. In general, the longitudinal modes associated with each cavity can be eliminated by using distributed Bragg gratings (DBG) [22] or periodic *PT* symmetric structures [17]. On the other hand, eliminating the transverse collective modes of the array is a daunting job. While several methods have been proposed to overcome this obstacle and regulate the functionalities of these devices [23-25], controlling their emission characteristics by using practical schemes remains an open problem.

In this work we propose a scheme for filtering the undesired transverse supermodes of laser arrays by using the concept of SUSY and we analyze their optical properties at the lasing threshold.

## 2. SINGLE MODE LASING IN SUPERSYMMETRIC LASER ARRAYS

Laser arrays are devices that consist of several interacting laser cavities [18-21]. In integrated optics platforms, these cavities are usually made of waveguides, ring resonators or photonic crystal cavities. In typical laser arrays, all the cavities are pumped equally with an external power source which results in multimode oscillations. Later, selective current injection was proposed as a means to favor the emission of only one mode [23]. Here we propose a different and more straightforward approach to achieve single transverse mode operation, i.e. by engineering the non-Hermiticity of the laser arrays to achieve mode selection through supersymmetry.

Figures 1 (a) and (b) depict two different realizations of the proposed structure in both optical waveguide and cavity platforms. In contrast to previous studies in laser arrays, our proposed device consists of an optical lattice $A_1$ and a spectrally engineered auxiliary array $A_2$ that serves as a non-Hermitian loss element to suppress the unwanted transverse modes. In order to explain the principle of operation, we assume that the main array $A_1$ is made of $N$ coupled identical cavities, each having the same linear mode. The coupling lifts the $N$-fold degeneracy and results in $N$ linear supermodes of different frequencies. These modes can contribute to the lasing process [21], and this multimode character is the main reason for the chaotic emission in laser arrays [19]. In order to achieve stable steady-state lasing, all but one of these $N$ linear modes must be eliminated (we will refer to them as $E$ modes) to allow only one remaining mode ($L$ mode) to participate in the lasing action. Thus, in order for the auxiliary array $A_2$ to achieve this required task, it should provide three functionalities: 1) It should increase the thresholds of the undesired $E$ modes of $A_1$, 2) it should exhibit minimum influence on the desired $L$ mode, and 3) It should not introduce other linear modes of lower threshold. While different optimization techniques might be invoked to achieve these goals, discrete supersymmetry (DSUSY) [4] provides a straightforward solution without the complications and constraints of numerical optimization.

More specifically, by applying the DSUSY prescription as described in appendix A, we design the auxiliary array $A_2$ such that it has $N$-1 linear modes that have the same frequencies as the $E$ modes of the main array $A_1$ (see Fig.2 (a)) but with stronger loss. By coupling these two arrays, for example, through their innermost cavities as shown in Fig.1, the *E* modes of $A_1$ and their counterparts of $A_2$ form *N*-1 doublet states as shown in Fig.2 (b), and more importantly, these doublets have a stronger loss than the desired *L* mode; the latter does not couple to the auxiliary array and forms a singlet state. This singlet state lases when the pump power reaches its threshold, which is below those of the doublet states.

## 3. LINEAR THRESHOLD ANALYSIS AND NUMERICAL SIMULATIONS

To justify our proposal given in the previous section, here we present the linear threshold analysis of SUSY arrays with numerical examples. A laser system is composed of certain geometries that exhibit loss, gain and feedback. In the absence of the gain, the complex eigenvalues of the system exist in the lower half of the complex plane, and any initial excitation will decay exponentially with time. As the gain is increased gradually, the eigenvalues are pulled toward the real axis. The linear threshold of each mode is defined as the gain value at which the eigenfrequency of this mode becomes real, which results in a steady-state laser oscillation. Once the linear thresholds of all possible lasing modes are considered, the value of the smallest one gives the actual lasing threshold, and its difference with the next smallest threshold is a good measure of how strong the gain needs to be increased before more than one lasing mode is excited. In other words, this difference gives an estimate of the range of gain value for single-mode operation. This treatment is a good approximation and has been widely used [21] because the laser is a linear system at the actual (first) threshold, at which the lasing intensity is zero. We note that if the nonlinearity is strong due to either spectral or spatial hole burning interactions between different lasing modes, the single-mode operation can be further extended [36].

Achieve single mode operation thus requires either to pull only one mode preferably to the real axis as the gain increases while leaving other modes intact, or to push the undesired modes further down in the complex eigenvalue space by adding additional loss channels, which can be described as "selective Q-spoiling." While techniques based on selective mode pumping follow the first approach, our proposal here highlights the second strategy by introducing extra optical loss in the auxiliary array, for example, by depositing metallic films on top of its constitute cavities, which convert light intensity into heat. An alternative route for obtaining similar effects is to use deep Bragg gratings [26] that introduce optical losses by coupling light efficiently to the continuum radiation without producing excessive heat.

In order to confirm the validity of our approach, we consider a concrete example of an optical array made of 5 cavities having the same resonant frequency $\omega_o$ and a uniform inter-cavity coupling coefficients $\tilde{\kappa}$. The auxiliary SUSY partner of this structure can be constructed by using DSUSY as described in appendix A, and the desired $L$ mode is chosen to be the fundamental mode with the highest frequency. Next we assume that intra-coupling between the main and auxiliary array is given by $\kappa'/\tilde{\kappa} = 0.2$ while the loss coefficient of the SUSY array is taken to be $\gamma_t = \gamma + \gamma_o$, where $\gamma = 0.06\tilde{\kappa}$ and $\gamma_o$ represent the original loss of the individual cavities as described in details in the next section. Fig. 3 (a) depicts the normalized complex eigenvalue spectrum of this composite structure as obtained from the exact discrete system (blue triangles) when the applied uniform gain value across both arrays is equal to $\gamma_o$. As expected, only the singlet eigenmode reaches lasing threshold (i.e., the real-axis of the complex eigenvalue plane), and the doublet states are pushed down away from the real axis, indicating their higher thresholds. This behavior, denoted schematically by the dashed line in Fig.2 (b), is in direct contrast with the spectral distribution of eigenmodes of laser arrays in the absence of the auxiliary non-Hermitian SUSY partner array where all the modes starts to lase at the same gain threshold. Evidently, our strategy succeeds in achieving single transverse mode lasing in a straightforward manner that does not require any special fabrication technology or complex nonuniform current injection schemes.

We further confirm these conclusions by investigating the temporal dynamics of proposed geometry under an initial arbitrary noise distribution as shown in Fig. 3 (b). Clearly, as time evolves, all the higher order optical eigenmodes suffer from dissipation and only the optical power obtained by projecting the initial noise distribution on the fundamental supermode survives. Note that only the main array $A_1$ is shown in Fig.3 (b) since the signal in the superpartner structure remains very weak during evolution.

As we have shown, both spectral analysis and temporal dynamics for the above example confirm single mode operation of laser arrays when a SUSY partner structure is introduced as an additional loss channel for the undesired $E$ modes without affecting the desired $L$ mode. Therefore, our approach can have a significant impact on both the fundamental science aspect and industry applications of laser arrays. Below we present the analytical results of the SUSY arrays and compare them with the numerical simulations presented above.

**4. ANALYTICAL RESULTS: SUPERMODE COUPLED EQUATIONS FOR SUSY ARRAYS**

In order to gain better understanding of the numerical linear threshold analysis presented above, we develop a supermode coupled mode theory (SCMT) that treats the interaction between the supermodes of both arrays rather than dealing with the evanescent coupling between individual cavities. In this context, we note that strength of the interaction between the supermodes is rather governed by their spatial overlap, which varies from one doublet to another.

We start our analysis by donating the eigenvectors of the main array by $\vec{V}_m = [v_{m,1} \; v_{m,2} \; ... \; v_{m,N-1} \; v_{m,N}]^T$ and those of its superpartner lattice by $\vec{U}_l = [u_{l,1} \; u_{l,2} \; ... \; u_{l,N-2} \; u_{l,N-1}]^T$, where the integers $m$ and $l$ take the values $1 \leq m \leq N$ and $1 \leq l \leq N-1$, respectively. $N$ here is the total number of optical cavities of the main lattice and the superscript $T$ denotes matrix transpose. If the spectrum of the SUSY array is constructed as described in appendix A where the fundamental optical mode does not have a superpartner mode, the optical tunneling can be assumed to occur only between $\vec{V}_{l+1}$ and $\vec{U}_l$. By denoting the coupling coefficients between the inner most cavities is given by $\kappa'$, we find that the overlap between the supermodes $\vec{V}_{l+1}$ and $\vec{U}_l$ is given by $\kappa_l = v_{l+1,N} \kappa' u_{l,N}$, where the eigenvectors have been normalized by $\sum_{m=1}^{N} v_{l+1,m}^2 = \sum_{m=1}^{N} u_{l,m}^2 = 1$. The SCMT between the degenerate eigenstates of both arrays (the ones that form a doublet) now takes the form:

$$i \frac{dV_{l+1}}{dt} - \omega_l V_{l+1} + \kappa_l U_l = 0$$
$$i \frac{dU_l}{dt} - \omega_l U_l + i\gamma U_l + \kappa_l V_{l+1} = 0$$
(1)

Here $V_{l+1}$ and $U_l$ are scalar quantities that represent the modal amplitude associated with the two eigenvectors $\vec{V}_{l+1}$ and $\vec{U}_l$, respectively while $\omega_l$ is their resonant frequency. We note that each of the $(2N-1)$ cavities in the array can support more than one $\omega_l$ in general, but here we focus on the single-mode case where only one of them is relevant, which can be achieved using distributed Bragg gratings (DBG) [22] or periodic *PT* symmetric structures [17]. In Eq. (1), we have also introduced the $\gamma$ term to account for a uniform and stronger optical loss in the superpartner lattice as described in section 2. The original homogenous loss of the individual cavities can be first neglected in our analysis, since it can be included simply as an imaginary part of $\omega_l$, which causes the same threshold increase of all modes. We will come back to this issue later when discussing the quantitative difference between the thresholds of the singlet state and the doublet states.

The eigenvalues associated with Eq. (1) that vary as $\exp(-i\Omega t)$ are given by $\Omega_l^\pm = \left(\omega_l \pm \sqrt{\kappa_l^2 - \left(\frac{\gamma}{2}\right)^2}\right) - \frac{\gamma}{2}i$. Clearly two distinct regimes can be identified for the eigenvalues depending on the ratio between $\kappa_l$ and $\gamma$. When $\kappa_l/\gamma > 0.5$, any two resonances belonging to the same doublet will have the same resonant lifetime $2/\gamma$ and their frequency split is given by $\Delta\Omega_l = \Omega_l^+ - \Omega_l^- = 2\sqrt{\kappa_l^2 - \left(\frac{\gamma}{2}\right)^2}$. In this regime, doublet states are formed as symmetric and anti-symmetric superposition of the supermodes of the two individual lattices, and hence both have similar intensity distributions. On the other hand, when $\kappa_l/\gamma < 0.5$, the two eigenmodes share the same resonant frequency while their resonance bandwidth ($img\{\Omega_l^\pm\}$) becomes different and their intensity distribution becomes strongly localized in either the main array or its superpartner structure. This phase transition is known as spontaneous *PT* symmetry breaking [27-35].

The onset of this transition at the point $\kappa_l/\gamma = 0.5$ is marked by an exceptional point [11, 12]. Note that according to the above model, the singlet eigenmode will always have zero loss. On the other hand, the loss factor of any of the doublets remains finite and approaches zero only in the limit when $\kappa_l/\gamma \ll 1$. This immediately suggests that the singlet supermode exhibits lower lasing threshold than any other eigenmode in the spectrum. Figure 3 (a) depicts the eigenvalue spectrum of the array considered in section two by using SCMT as shown by the red dots. Evidently good agreement between the exact diagonalization and the SCMT is observed in most cases. Note also that SCMT predicts the nonuniform splitting of the doublets. However, some small discrepancies between both calculations do exist as indicated by the closed dashed curves. We show in Appendix B that this is an outcome of non-resonant interactions using the Brillouin-Wigner (BW) perturbation method [38].

These results thus quantify and confirm the possibility of achieving lasing action only in the singlet state by applying a uniform optical gain to all the cavities. To calculate the lowest threshold of the doublet states in comparison with the singlet, we denote the initial material and radiation loss in every individual cavity in both lattices by $\gamma_0$. By adding the extra loss $\gamma$ to the superpartner structure as described in section 2, its total loss becomes $\gamma + \gamma_0$. Evidently, $\gamma_0$ increases the thresholds of all modes in the SUSY array by the same amount. The singlet state reaches threshold when the applied gain, denoted by $g_s$, equals $\gamma_0$. The lower threshold for a pair of doublet states $\Omega_l^\pm$, denoted by $g_l$ and calculated using SCMT in the absence of nonlinear interactions, is given by $g_l = \gamma_0 + \frac{\gamma}{2} - \sqrt{\left(\frac{\gamma}{2}\right)^2 - \kappa_l^2}$ in the *PT* broken phase and $g_l = \gamma_0 + \frac{\gamma}{2}$ in the *PT* symmetric phase. For a given structure (and hence $\kappa_l$), clearly $g_l(\gamma)$ reaches its maximum value right at the EP when $\gamma = 2\kappa_l \equiv \gamma^{EP}$. Under these favorable conditions for suppressing the doublet modes from lasing, the maximum ratio of the lowest doublet threshold and the singlet threshold is given by $1 + \frac{\kappa_l}{\gamma_0}$. From this analysis, it is clear that a strong coupling $\kappa_l$ at the EPs and a low loss $\gamma_0$ in the main array facilitates a significant single-mode lasing action. As we noted before, using this linear analysis to predict lasing threshold is a well established technique and has been verified before in literature [15, 21]. However, in order to simulate emission dynamics beyond the lowest lasing threshold, one has to resort to more complicated nonlinear models. Nevertheless, it is reasonable to expect that in a SUSY array the saturated gain medium in the main lattice will further suppress the doublet states from lasing [36].

## 5. DISCUSSION AND CONCLUSION

The SCMT presented above provides a consistent picture with our physical intuition. For certain design parameters, however, exceptions take place where the lasing mode having the lowest threshold turns out to be a doublet state. This finding is exemplified in Fig. 4 where the singlet state has the second highest frequency. As we show in Appendix B using the Brillouin-Wigner perturbation method [38], these rare exceptions are due to nonresonant interactions between multiple supermodes in the main array and the auxiliary array, a feature that was previously overlooked in the study of supersymmetric optical structures.

In presenting the scheme and analysis of SUSY arrays above, we have assumed a uniformly applied gain in both the main and auxiliary arrays while introduced a stronger loss in the latter. Another alternative to favor lasing in the singlet state is to uniformly pump only the main array, without the need to introduce additional loss to the auxiliary array. In this case, the threshold of the singlet state remains the same at $g_s = \gamma_0$ since it does not couple to the superpartner lattice. On the other hand, the eigenfrequencies of the doublet states as obtained by using the SCMT are now given by $\Omega_l^\pm = \left(\omega_l - i\gamma_0 \pm \sqrt{\kappa_l^2 - \left(\frac{g}{2}\right)^2}\right) + \frac{g}{2}i$. Clearly, when $\gamma_0 < \kappa_l$ the doublet reaches their lasing threshold at $g_l = 2\gamma_0$ and remains in the $PT$ symmetric phase. Otherwise (i.e., when $\gamma_0 > \kappa_l$), the pair start to lase in the $PT$ broken phase and their lower threshold is given by $g_l = \gamma_0 + \frac{\kappa_l^2}{\gamma_0} < 2\gamma_0$ [15]. Thus for a given loss coefficient $\gamma_0$ in every individual cavity of both arrays, operation in the $PT$ symmetric phase is more favorable to suppress lasing in the doublet states, achievable by making $\kappa_l$ greater than $\gamma_0$. Within this scenario, the maximum ratio of the lower doublet threshold and the singlet threshold in the linear analysis is then exactly 2.

In conclusion, we have introduced the concept of SUSY laser arrays that are capable of supporting laser oscillations in the singlet states only. If each cavity of the SUSY array supports only a single resonant frequency [17, 22], then single-mode lasing is possible in the corresponding singlet state. We have also shown that under certain operation conditions, anomalous lasing can occur where one of the doublet eigenmodes exhibits lower lasing threshold than the singlet state due to non-resonant interactions.

# APPENDIX A: CONSTRUCTION OF SUPERSYMMETRIC PARTNER ARRAY

In order to construct the supersymmetric auxiliary lattice, we note that the Hamiltonian of the main array $A_1$ can be described by a discrete $N \times N$ tri-diagonal Hamiltonian matrix $H^{(1)}$ whose elements are given by $H^{(1)}_{n,n} = \omega_o$ and $H^{(1)}_{n,n+1} = H^{(1)}_{n+1,n} = \tilde{\kappa}$. The Hamiltonian of its superpartner structure that does not contain the $m^{\text{th}}$ mode of $A_1$ can be constructed through discrete SUSY transformation [4,5], and it is given by $H^{(2)} = (RQ + \omega_m I)_{(N-1)} = (Q^T H^{(1)} Q)_{(N-1)}$, where $\omega_m$ is the eigenfrequency of the $m^{\text{th}}$ mode of $A_1$ and $I$ is the unit matrix of dimensions $N \times N$. The subscript indicates that $H^{(2)}$ is constructed by selecting only the upper-left block diagonal matrix of dimensions $(N-1) \times (N-1)$ of the larger matrix in the parentheses after isolating the zero $m^{\text{th}}$ eigenvalue at the last raw and column. Here $Q$ and $R$ are the QR factorization matrices of $H^{(1)} - \omega_m I$ [40].

# Appendix B: BRILLOUIN-WIGNER PERTURBATION ANALYSIS FOR SUSY ARRAYS AND NONRESONANT INTERACTIONS

In order to understand the difference of the numerical simulation and SCMT in Fig. 3(a), and more importantly, the change of lasing order briefly mentioned in the conclusion section of the main text, we analyze the SUSY array using a perturbation theory. We start by writing the total Hamiltonian of the system in the form:

$$H_{tot} = H_o + H_I,$$

$$H_o = \begin{bmatrix} H_1 & 0 \\ 0 & H_2 \end{bmatrix} \quad \& \quad H_I = \begin{bmatrix} 0 & Z \\ Z^T & 0 \end{bmatrix}$$

(2)

In Eq. (2), $H_o$ is the unperturbed Hamiltonian and is part of a closed algebra [2] while $H_I$ is a Hermitian perturbation that couples the main array $A_1$ to its superpartner lattice $A_2$. $Z$ is in general a $N \times (N-1)$ matrix ($N = 5$ in the examples given in the main text) with $z_{N,1} = \kappa'$ and zero entries otherwise. In principle, one can apply the usual Rayleigh-Schrodinger (RS) perturbation theory [37] to study the eigenvalues and eigenmodes of $H_{tot}$. However, the procedure is complicated by the multiple two-fold degeneracies associated with $H_o$. A powerful alternative is to employ Brillouin-Wigner (BW) perturbation method [38]. While both RS and BW perturbation methods agree to first order, BW method offers more accurate results for higher orders calculations with the need for any special treatments for degenerate eigenstates. These advantages come at the expense of solving polynomial equations in order to obtain the perturbed eigenvalues. In particular, the expression for the new eigenvalues using BW method takes the form:

$$\Omega_l^{new} = \Omega_l + \sum_{j=1}^{\infty} S_j,$$

$$S_1 = \vec{W}_l^T H_I \vec{W}_l$$

$$S_2 = \sum_{m \neq l} \frac{(\vec{W}_l^T H_I \vec{W}_m)(\vec{W}_m^T H_I \vec{W}_l)}{\Omega_l^{new} - \Omega_m}$$

$$\vdots$$

$$S_{j+1} = \sum_{m_1 \neq l} \sum_{m_2 \neq l} \cdots \sum_{m_j \neq l} \frac{(\vec{W}_l^T H_I \vec{W}_{m_1})(\vec{W}_{m_1}^T H_I \vec{W}_{m_2}) \cdots (\vec{W}_{m_j}^T H_I \vec{W}_l)}{(\Omega_l^{new} - \Omega_{m_1})(\Omega_l^{new} - \Omega_{m_2}) \cdots (\Omega_l^{new} - \Omega_{m_j})}$$

, (3)

where $\Omega_m$ and $\vec{W}_m$ are the unperturbed eigenvalues and eigenvectors of $H_o$, respectively and $\Omega_l^{new}$ is the new perturbed eigenfrequency associated with the $l^{th}$ mode.

The subscripts $m_i$ in the summation for the $S_j$ terms run over all the modes of the systems except the indicated ones. As we have noted, the spectrum of $H_o$ contains multiple double degeneracies. Thus the eigenvector bases are not unique. Here we use the bases $\begin{bmatrix} \vec{V}_{l_1} \\ \vec{0}_{4\times 1} \end{bmatrix}$ and $\begin{bmatrix} \vec{0}_{5\times 1} \\ \vec{U}_{l_2} \end{bmatrix}$ where the subscripts $l_{1,2}$ run over all the modes of $H_{1,2}$, respectively, and $\vec{V}_{l_1}$ & $\vec{U}_{l_2}$ are their associated eigenvectors. Here $\vec{0}_{n\times 1}$ is a column vector of dimensions $n\times 1$. In these bases, the first order correction $T_1$ in the above formula is zero for our problem. Interestingly, if we restrict our perturbation expansion to second order approximation and we retain only the resonant term in the summation of $T_2$ (i.e. the term that satisfies $\text{Re}\{\Omega_l - \Omega_m\} = 0$), Eq.(3) reduces to the SMCT and the two eigenfrequencies $\Omega_l^{\pm}$ described above emerge naturally as solutions of a quadratic equation. However, in order to proceed beyond the SMCT, one has to consider the full polynomial equation with its all possible solutions. These solutions can be found graphically or by using any of the well-developed numerical techniques. Among this family of solutions, only those that represent relatively small perturbation over the unperturbed eigenmode should be retained while the others must be discarded.

A simpler procedure for finding the relevant solutions can be obtained by noting that in our particular SUSY configuration, the strongest contributions to the expansion (3) arise from the interaction between resonant modes. By neglecting terms higher than $T_2$ and substituting $\Omega_l^{new} = \Omega_l$ in every term in the right hand side of $T_2$ except the resonant one, we arrive at a quadratic equation whose two solutions are given by:

$$\Omega_l^{new} = \frac{\Omega_l + \Omega_n + \delta}{2} \pm \sqrt{\left|\left(\vec{W}_n^T H_I \vec{W}_l\right)\right|^2 + \left(\frac{\Omega_l + \delta - \Omega_n}{2}\right)^2} \quad , \quad (4)$$

where the complex parameter $\delta = \sum_{m \neq l,n} \frac{\left|\left(\vec{W}_m^T H_I \vec{W}_l\right)\right|^2}{\Omega_l - \Omega_m}$ characterizes the second order interaction between non-resonant SUSY eigenstates while the resonant eigenvalue $\Omega_n$ satisfies the relation $\text{Re}\{\Omega_l - \Omega_n\} = 0$. We verify formula (4) by first revisiting the example of Fig.3. Recall that in contradiction with the exact diagonalization of the full discrete Hamiltonian, SCMT did not account for *PT* spontaneous symmetry breaking of some of the doublets in the spectrum as highlighted by the dashed closed curves in Fig.3 (a). Equation (4), on the other hand, correctly predicts the onset of *PT* phase transition in both cases. Finally by applying the BW perturbation analysis of Eq. (4) to the example associated with Fig.4, we find that BW analysis remarkably reproduce the unexpectedly anomalous spectrum with high accuracy. It is thus clear that the counter-intuitive shuffling of the orders of the lasing modes indicated schematically in Fig.4 (a) is a direct outcome of the non-resonant interactions between the modes [39].

**Figure captions**

Figure 1 (Color online) A schematic of possible physical realizations of supersymmetric laser arrays using (a) waveguides platform and (b) optical resonators configurations.

Figure 2 (Color online) Conceptual demonstration of singlet lasing using a supersymmetric array. (a) A schematic of the individual spectra of an optical array and its superpartner that does not share the singlet state (with highest frequency). (b) Shows the spectrum of the combined system. It consists of a set doublets and one singlet. When optical loss is added to the superpartner lattice, supermode coupled mode theory (SMCT) predicts that the singlet state (dashed line) will exhibit the lowest lasing threshold.

Figure 3 (Color online) (a) Eigenvalue spectrum of a supersymmetric array when $\kappa'/\tilde{\kappa} = 0.2$ and $\gamma/\tilde{\kappa} = 0.06$ where $\tilde{\kappa}$ is the uniform coupling coefficients between the cavities of the original array, $\kappa'$ is the coupling constant between the inner most channels of the main lattice and its superpartner while $\gamma$ is the uniform optical loss coefficient of the superpartner structure $A_2$. Results obtained using exact diagonalization of the discrete system and from supermode coupled mode theory (SCMT) are compared on the same figure. (b) Temporal evolution of an initial arbitrary optical intensity distribution in this SUSY array. As expected, only the fundamental eigenmode of $A_1$ survives as time lapses. Only the optical cavities of the array $A_1$ are depicted.

Figure 4 (Color online) (a) Eigenvalues distribution of a supersymmetric array made of 9 resonators (5 of which belong to $A_1$) designed to eliminate the $4^{th}$ order eigenmode of $A_1$ from the spectrum of $A_2$. In this example $\kappa'/\tilde{\kappa} = 0.2$ and $\gamma/\tilde{\kappa} = 0.5$. The doublet state indicated by the dashed line reaches the lasing point before the singlet state as one would have expected. This counter-intuitive result is confirmed in (b) where an initial state made of an equal superposition of the 5 eigenmodes of $A_1$ (middle panel) evolves to the doublet eigenstates (upper panel) instead of the singlet (lower panel). Panel (c) presents the time dependent dynamics for the projection coefficients $C_n(t)$, defined by $\vec{\varphi}(t) = \sum_{n=1}^{N} C_n(t) \vec{V}_n$ in the main array. $C_n(t=0) = 1/\sqrt{5}$ for all $n$.

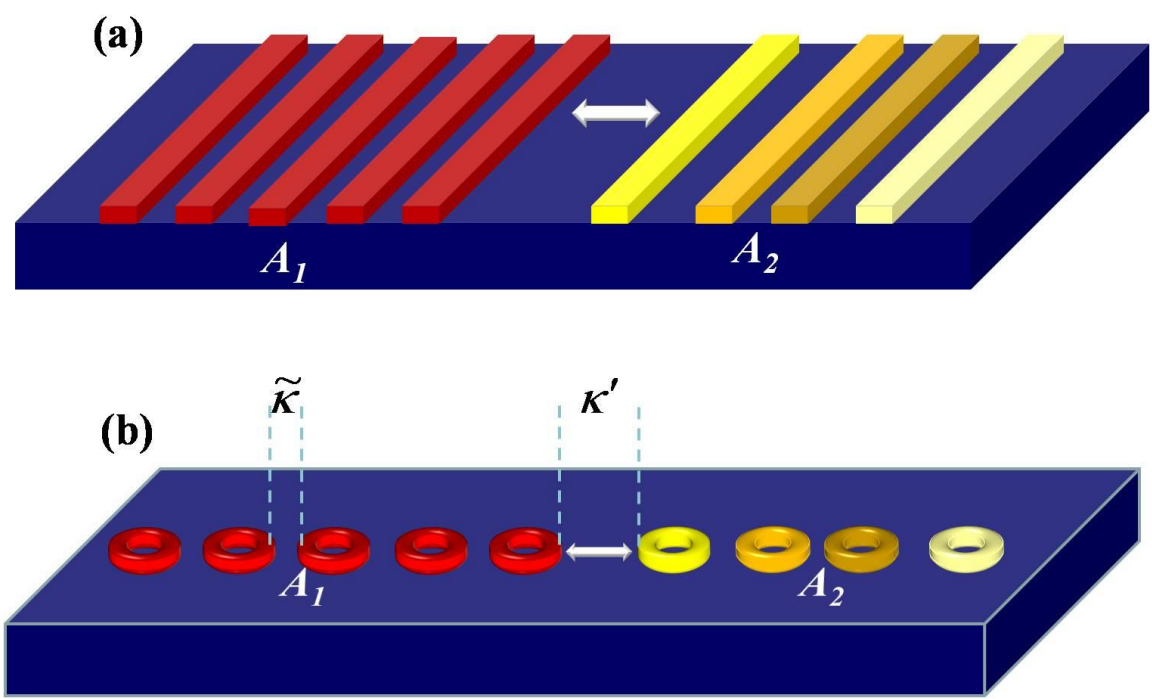

Fig.1

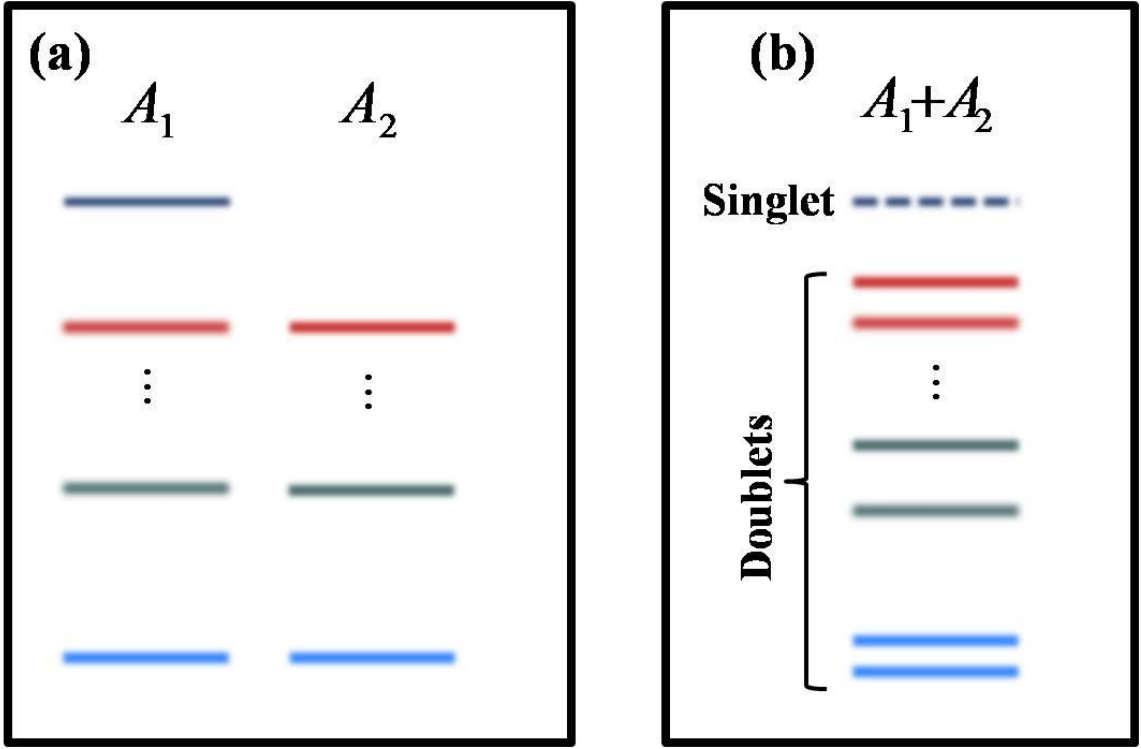

Fig.2

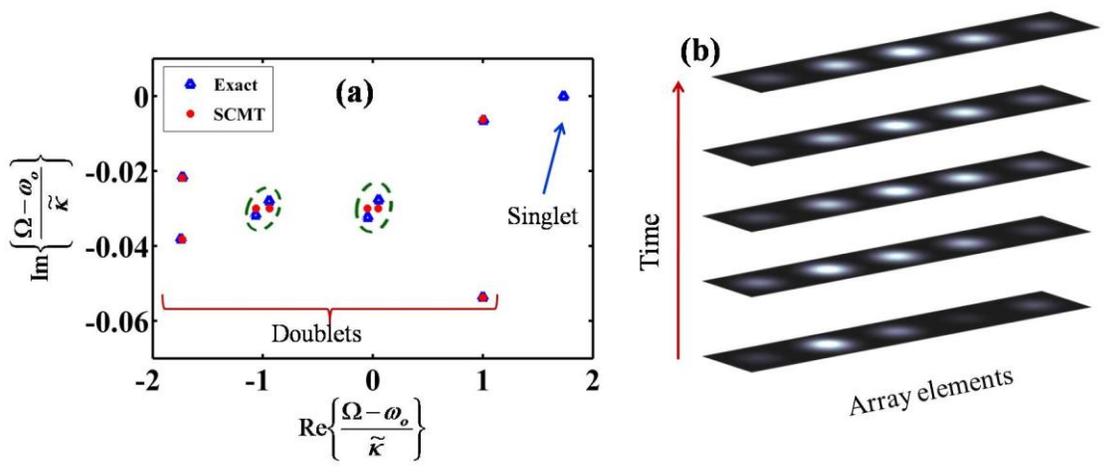

Fig.3

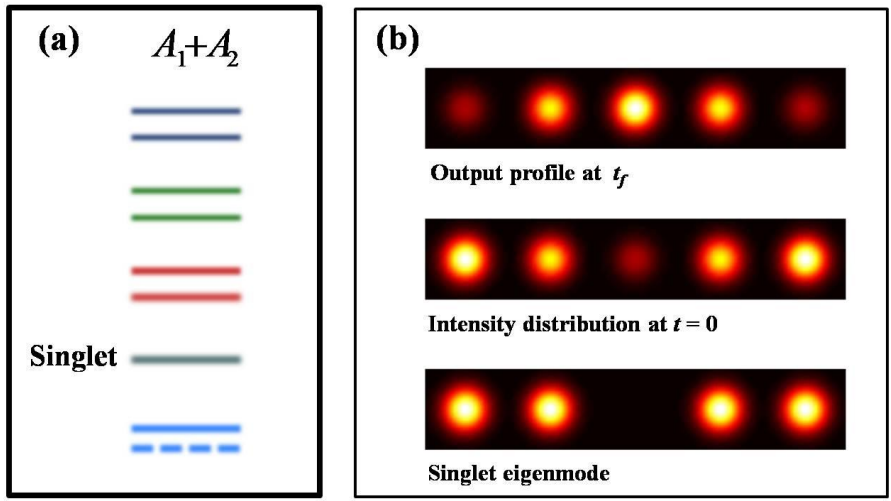
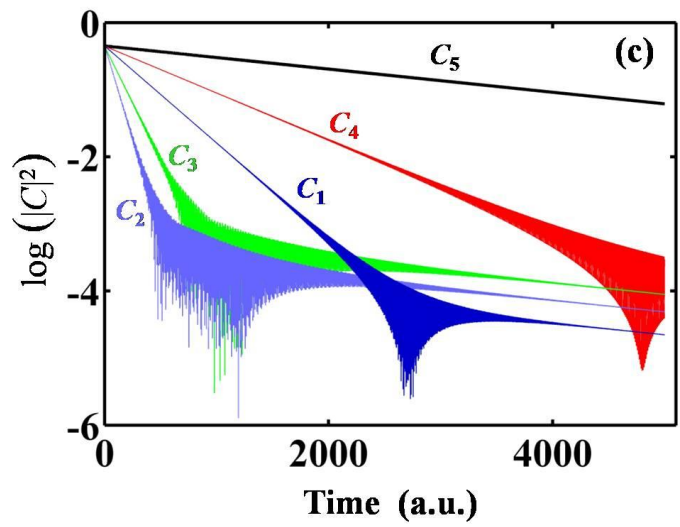

Fig.4